\documentclass[aps,prc,reprint]{revtex4-1}
\usepackage{amssymb,amsbsy,amsfonts,amsmath}
\usepackage{epsfig,color}
\usepackage{epstopdf}
\usepackage[citecolor=blue,colorlinks=true,linkcolor=blue,urlcolor=blue]{hyperref}
\begin{document}
\title{Phase-dependent cross sections of deuteron-triton fusion in dichromatic intense fields with high-frequency limit}
\author{Wenjuan Lv$^{1}$}
\author{Hao Duan$^{2,3}$}
\email{duan$_$hao@iapcm.ac.cn}
\author{Jie Liu$^{1,4}$}
\email{jliu@gscaep.ac.cn}
\affiliation{$^{1}$Graduate School, China Academy of Engineering Physics, Beijing 100193, China}
\affiliation{$^{2}$Laboratory of Computational Physics, Institute of Applied Physics and Computational Mathematics, Beijing 100088, China}
\affiliation{$^{3}$Institute of Applied Physics and Computational Mathematics, Beijing 100088, China}
\affiliation{$^{4}$CAPT, HEDPS, and IFSA Collaborative Innovation Center of MoE, Peking University, Beijing 100871, China}

\begin{abstract}
We investigate the influence of strong dichromatic laser fields(i.e. $1\omega-2\omega$ and $1\omega-3\omega$) with
high-frequency limit on the cross sections of deuteron-triton(DT) fusion in Kramers-Henneberger(KH) frame.
We focus on the transitions of phase-dependent effects depending on a dimensionless quantity $n_{d}$, which equals the ratio of the quiver oscillation amplitude to the geometrical touching radius of the deuteron and triton as defined in our previous research. 
Theoretical calculations show that the angle-dependent as well as phase-dependent Coulomb barrier penetrabilities can be enhanced in dichromatic intense fields, and the corresponding angle-averaged penetrabilities and the fusion cross sections increase significantly compared with field-free case.
Moreover, we find that there are twice shifts of the peak in the cross sections whenever the frequency becomes sufficiently low or the intensity sufficiently high. The reason for the first shift is the angle-dependence effects for sub-barrier fusion, while the second shift is due to the accumulation of over-barrier fusion, these mechanisms are analyzed in detail in this paper.

\end{abstract}
\maketitle
\section{Introduction}
The application of strong laser technology in atomic physics and nuclear physics has attracted more and more attentions. Intense lasers can be applied to atomic ionization \cite{Joachain,Liu,Mima} , charged-particle acceleration \cite{Mangles,Geddes,Faure} and also provide a new way for manipulating nuclear processes.

It is found that intense lasers can induce resonance internal conversion, i.e., F. F. Karpeshin considered mechanisms of isomer pumping via a laser-radiation-induced resonance conversion and of isomer energy triggering in a resonance laser radiation field \cite{Karpeshin}. And intense lasers can accelerate nuclear fission processes \cite{Bai201801,Qi2020}, especially in the aspect of increasing $\alpha$-decay rates \cite{Delion2017,Bai201802,Qi2019} by modifying the Coulomb potential barrier. However, there is no conclusive result on whether laser-induced enhancement of the decay rate is considerable or not because there is no successful experiment to prove these theories \cite{Palffy, Ghinescu}.
In particular, since nuclear fusion processes are mainly associated with light nuclei, laser manipulation will be more effective because of the relatively large charge-mass ratio compared with that in the heavy nuclei processes. Friedemann Queisser and Ralf Sch$\ddot{u}$tzhold studied whether the tunneling probability could be enhanced by an additional x-ray free electron laser (XFEL). They find that the XFEL frequencies and field strengths required for this dynamical assistance mechanism should come within reach of present-day or near-future technology. Apart from the deformation of the potential barrier, the time dependence plays a crucial role for assisting tunneling through the Coulomb barrier \cite{Friedemann2019}.

Based on these, we investigated deuteron-triton(DT) fusion cross sections in the presence of electromagnetic fields with high intensity and high frequency \cite{Lv2019}. With the help of the Kramers-Henneberger(KH) transformation \cite{Henneberger}, we have shown that the corresponding Coulomb barrier penetrabilities increase significantly due to the depression of the time-averaged potential barrier. As a result, we have found that DT fusion cross sections can be enhanced depending effectively on a dimensionless quantity $n_{d}$, which equals the ratio of the quiver oscillation amplitude to the geometrical touching radius of the deuteron and triton. 
Wang have shown that intense low-frequency laser fields, such as those in the near-infrared regime for the majority of intense laser facilities around the world, are highly effective in transferring energy to the DT system and enhancing the fusion probabilities \cite{Wang2020}.

The above researches are focus on the effects of intense monochromatic laser fields on nuclear fusion processes. Similarly, dichromatic laser fields display important novel features that cannot be seen with monochromatic laser fields driving. Especially the phase-dependent effects, which have been widely studied in the ionization of atoms \cite{Schafer 1992, Chen Jing 1998, Chen Jing 2000, Cheng Taiwang 1999, Cheng Taiwang 2000 1, Cheng Taiwang 2000 2, Cheng Taiwang 2002}.

%
%
%
%
%
%

Our goal is to study DT fusion cross sections in the presence of high intensity dichromatic laser fields (i.e. $1\omega-2\omega$ and $1\omega-3\omega$) with high-frequency limit, especially the transitions of phase-dependent effects.

The paper is organized as follows. We introduce our model in Sec. II, the shape of the time-dependent dichromatic laser fields and the quiver oscillations of DT system in the presence of these fields are shown in this part. We provide the time-averaged two-body interaction potential between the DT system, including the short-range attractive nuclear potential and long-range repulsive Coulomb potential in Sec. III. Our results on the Coulomb barrier penetrabilities and DT fusion cross sections are given in Sec. IV where we show the twice shifts of the peak in the cross sections and analyse the reason combined with the potential in Sec. III. Sec. V presents our conclusion.

\section{Model}
Nuclear fusion is commonly believed to consist of three processes \cite{Atzeni,Gamow,Bosch,Lv2019}.
In the presence of dichromatic laser fields (i.e. $1\omega-2\omega$), The time-dependent Schr\"{o}dinger equation in KH frame under the condition of nonrelativistic dipole approximation is
\begin{eqnarray}
i\hbar\frac{\partial}{\partial t}\Psi_{\mathrm{kh}}\left(t,\vec{r}_{\mathrm{kh}}\right)&=&
\left(\frac{\vec{p}^{2}_{\mathrm{kh}}}{2m}+V_{\mathrm{kh}}
\left(t,\vec{r}_{\mathrm{kh}}\right)\right)\Psi_{\mathrm{kh}}
\left(t,\vec{r}_{\mathrm{kh}}\right).\nonumber\\
\label{TDSE}
\end{eqnarray}
\begin{figure}[!b]
\centering
\includegraphics[width=\linewidth]{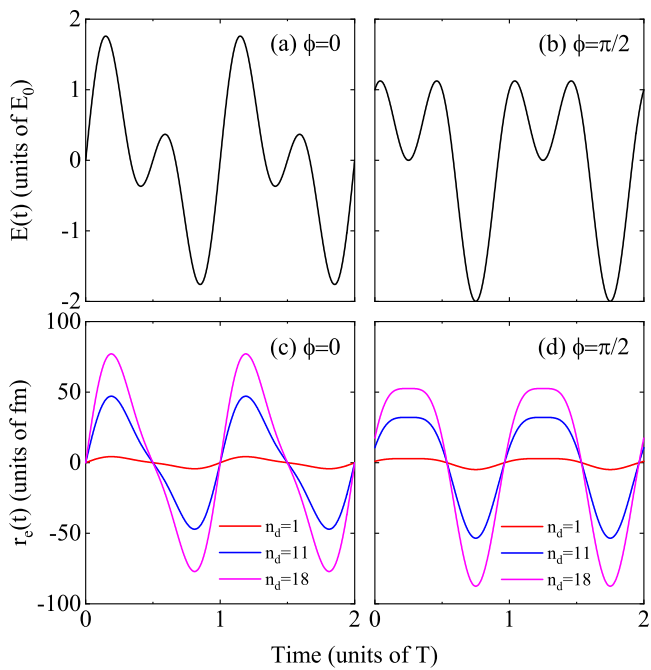}
\caption{(Color online) The time-dependent dichromatic electric fields (in units of $E_{0}$) for (a)$\phi=0$ and (b)$\phi=\pi/2$. (c)$\phi=0$ and (d)$\pi/2$ are the quiver oscillations of DT system in the fields for $n_{\mathrm{d}}=1,11$ and $18$, respectively.}
\label{figure1}
\end{figure}
Here, $\Psi_{\mathrm{kh}}$ is the wavefunction under the KH frame. The coordinate operator is $\vec{r}_{\mathrm{kh}}(t)=\vec{r}-\vec{r}_{\mathrm{e}}(t)$, where $\vec{r}$ is relative displacement vector of deuteron and triton, and $\vec{r}_{\mathrm{e}}(t)$ is the quiver motion of a free nucleus in the dichromatic laser fields. Supposing that the fields are a laser field and its second harmonic, which are linearly polarized along the $z$-axis, i.e., $\vec{E}(t)=\hat{e}_{z}E_{0}\left(\sin\omega t+\sin\left(2\omega t+\phi\right)\right)$, then we can get $\vec{r}_{\mathrm{e}}(t)=\hat{e}_{z}r_{\mathrm{e}}\left(\sin\omega t+\sin\left(2\omega t+\phi\right)/4\right)$ with $r_{\mathrm{e}}=\mathrm{e}\sqrt{2c\mu_{0}I}/5m\omega^{2}$, where  $m=m_{1}m_{2}/(m_{1}+m_{2})$ is the reduced mass and $\phi$ is the relative phase. Let us introduce a dimensionless quantity $n_{\mathrm{d}}=r_{\mathrm{e}}/r_{\mathrm{n}}=4.89\times10^{-6}\sqrt{I}/(\hbar\omega)^{2}$, here, the geometrical touching radius $r_{\mathrm{n}}=1.44(A^{1/3}_{1}+A^{1/3}_{2})\ \mathrm{fm}$, the units of $I$ and $\hbar\omega$ are $\mathrm{W/cm^{2}}$ and eV, respectively.
The ratio of $r_{\mathrm{e}}$ to $r_{\mathrm{n}}$ determines how external electromagnetic fields manipulate nuclei collision processes.
The position and momentum operators maintain the commutation relation $[r^{i}_{\mathrm{kh}}(t),p^{j}_{\mathrm{kh}}(t)]=i\hbar\delta^{i,j}$.
The expression of time-dependent potential is the same as that in our previous paper \cite{Lv2019}.
%

For $\phi=0$ and $\pi/2$, we show the shape of two cycles of the time-dependent dichromatic electric fields and the quiver oscillations of DT system in these fields for $n_{\mathrm{d}}=1,11$ and $18$ in Fig. \ref{figure1}. It can be seen from Fig. \ref{figure1} (a) and (b) that the maximum instantaneous field is the least when $\phi=0$, this maximum value of $1.76E_{0}$ is obtained twice per cycle, while the maximum instantaneous field is the greatest when $\phi=\pi/2$, this maximum value of $2E_{0}$ is obtained once per cycle. Fig. \ref{figure1} (c) and (d) are the quiver oscillations accordingly, and the range are about $[-1.10n_{\mathrm{d}}, 1.10n_{\mathrm{d}}]$ for $\phi=0$ and $[-1.25n_{\mathrm{d}}, 0.75n_{\mathrm{d}}]$ for $\phi=\pi/2$, respectively.
And according to our previous study \cite{Lv2019}, the corresponding laser parameters for $n_{\mathrm{d}}=1$, $11$ and $18$ are $2.80\times10^{3}\ \mathrm{eV},\ 2.57\times10^{24}\ \mathrm{W/cm^{2}}$; $2.80\times10^{3}\ \mathrm{eV},\ 3.16\times10^{26}\ \mathrm{W/cm^{2}}$; and $2.80\times10^{3}\ \mathrm{eV},\ 8.36\times10^{26}\ \mathrm{W/cm^{2}}$, respectively.




\begin{figure*}[!tb]
\centering
\includegraphics[width=\linewidth]{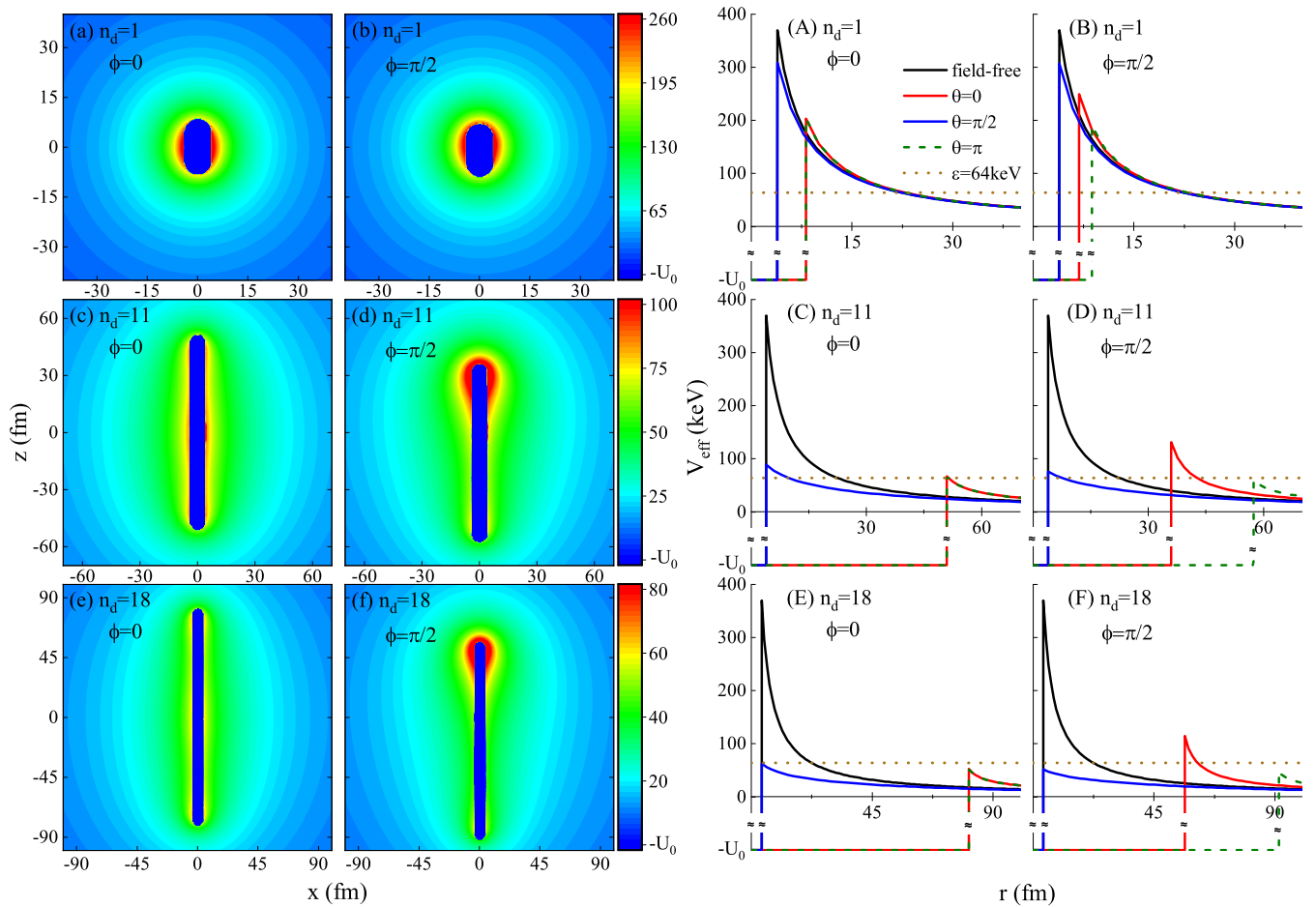}
\caption{(Color online) For $\phi=0$ and $\pi/2$, (a) to (f) contour plots on x-z section ($y=0$) of the effective potential for $n_{\mathrm{d}}=1$, $11$ and $18$, respectively. The blue areas represent the section of inner region $\mathrm{D_{in}}$ where the potential value is approximately $-U_0$. (A) to (F) $V_{\mathrm{eff}}$ for different inclination angles $\theta$ with respect to varied  $n_{\mathrm{d}}$.}
\label{figure2}
\end{figure*}

\section{Time-Averaged Potential}
Eq. (\ref{TDSE}) indicates that all time dependence of the problem is shifted into the potential function, and $V_{\mathrm{kh}}\left(t,\vec{r}_{\mathrm{kh}}\right)$ is just a two-body Coulomb potential dressed by a time-dependent harmonic oscillation origin $\vec{r}_{\mathrm{e}}(t)$ along the polarization direction $\hat{e}_{z}$ with a quiver oscillation amplitude $r_{\mathrm{e}}$. In this case, the space can be divided into two parts: the inner region denoted by a capsule-like region swept by the nuclear potential well $U_{0}$, which is approximately $30$ to $40$ MeV, i.e., $\mathrm{D_{in}}=\{\vec{r} |r_{\mathrm{kh}}(t)\leq r_{\mathrm{in}},\exists t\in[0,2\pi/\omega)\}$, where $r_{\mathrm{in}}$ is the boundary of capsule-like region. And the outer region denoted by $\mathrm{D_{out}}=\mathrm{R}^{3}/\mathrm{D_{in}}$. According to our previous study \cite{Lv2019}, when the characteristic collision time is much longer than the fields period, i.e., the time-dependent operator $\vec{r}_{\mathrm{kh}}(t)$ can be well approximated by its time-averaged value $\overline{\vec{r}_{\mathrm{kh}}(t)}=\vec{r}$ in high-frequency laser fields, indicating that the incident nucleus feels a time-averaged potential $\overline{V_{\mathrm{kh}}(t,\vec{r})}=V_{\mathrm{eff}}(\vec{r})$. We choose $1$ keV as the threshold of the laser frequency, beyond which the average scheme is valid.

The time-averaged potential in both its peak value and tunneling width is distorted in the presence of strong fields, and the effective potential, including the short-range attractive nuclear potential and long-range repulsive Coulomb potential, is dependent on $\phi$ and $\theta$. The corresponding time-averaged potential $V_{\mathrm{eff}}$ for $n_{\mathrm{d}}=1$, $11$, and $18$ are shown in Fig. \ref{figure2}, respectively. Along the polarization direction $\hat{e}_{z}$ of $\theta=0$, $\pi$, both the peak value and the barrier width are found to decrease significantly.
For $\phi=0$, The angle dependence of time-averaged potential is symmetrical for $\theta=\pi/2$. For $\phi=\pi/2$, such symmetrical structure doesn't exist.

\begin{figure}[!t]
\centering
\includegraphics[width=\linewidth]{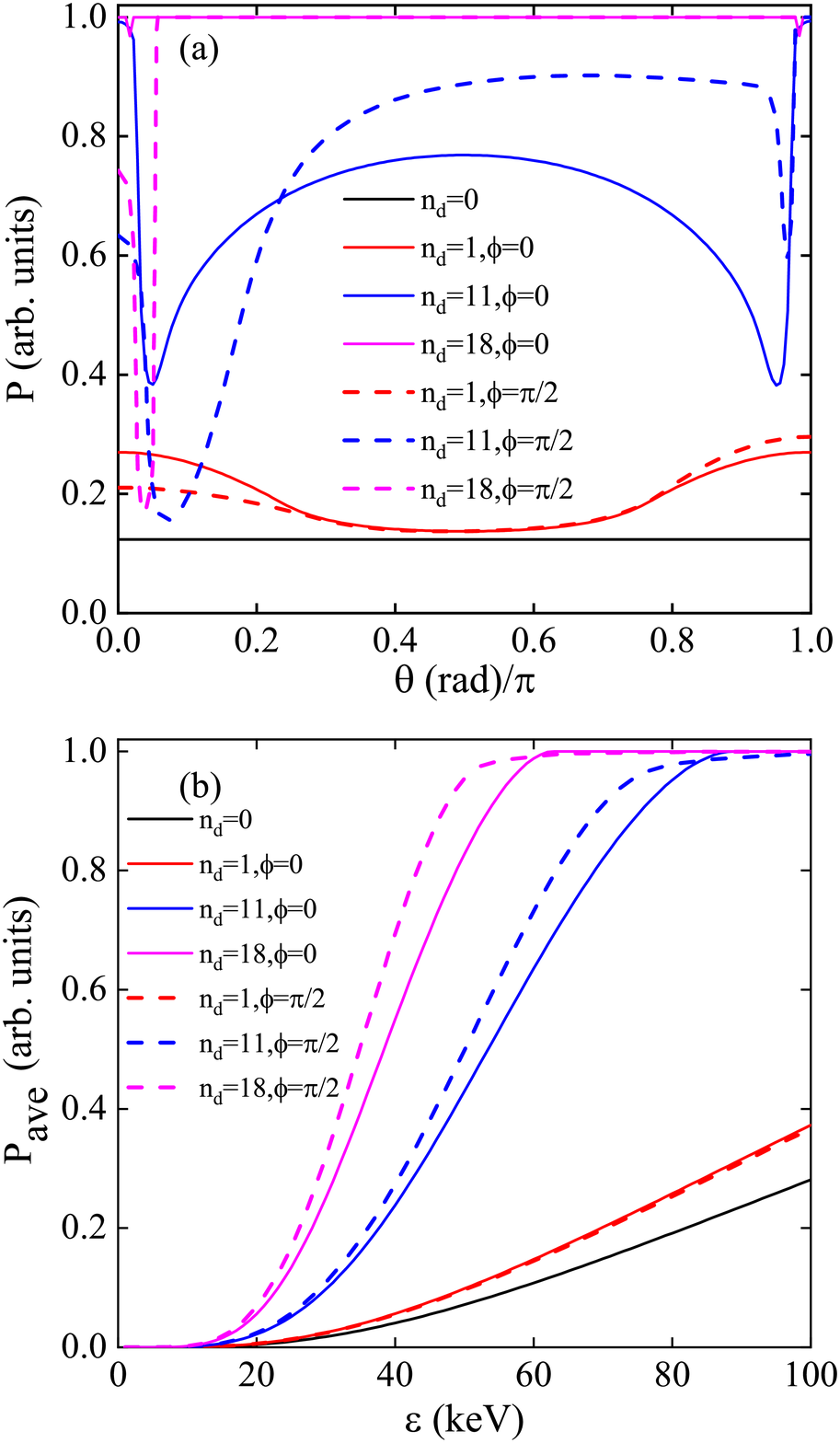}
\caption{(Color online) For $n_{\mathrm{d}}=0$, $1$, $11$ and $18$, (a) angle-dependent penetrability for collision energy of $\varepsilon=64$ keV. (b) angle-averaged penetrability versus collision energy.}
\label{figure3}
\end{figure}

\section{Main Results}
\subsection{Penetrability}
Using the Wenzel-Kramers-Brillouin (WKB) approximation \cite{Landau}, the penetrability through the Coulomb barrier at the collision energy $\varepsilon$ along the direction $\hat{r}$ can be given by
\begin{eqnarray}
P\left(\theta;\phi,\varepsilon,n_{\mathrm{d}}\right)=
e^{-\frac{2}{\hbar}\int_{r_{\mathrm{in}}}^{r_{\mathrm{out}}}\sqrt{2m\left(V_{\mathrm{eff}}
\left(\theta,r;\phi,n_{\mathrm{d}}\right)-\varepsilon\right)}dr},
\label{P}
\end{eqnarray}
where $r_{\mathrm{in}}$ and $r_{\mathrm{out}}$ are the inner and outer turning points, respectively. Due to the symmetry of the Hamiltonian, the penetrability is independent of the azimuth $\varphi$. The penetrability explicitly depends on the inclination angle $\theta$, the relative phase $\phi$, the incident energy $\varepsilon$ and the dimensionless parameter $n_{\mathrm{d}}$.

The angle-dependent penetrability can be readily obtained by numerically calculating Eq. (\ref{P}), and the results are plotted in Fig. \ref{figure3}. As shown in Fig. \ref{figure3} (a), the angle-dependent penetrability exhibit different structures for different laser parameters. For $\phi=0$, The angle-dependent penetrabilities are symmetrical for $\theta=\pi/2$ where we can get a minimum penetrability for $n_{\mathrm{d}}=1$ and a maximum penetrability for $n_{\mathrm{d}}=11$. In particular, the angle-dependent penetrabilities exhibit an interesting double-hollow structure for $n_{\mathrm{d}}=11$, this indicates that the penetrability can reach maxima in the directions parallel and perpendicular to the laser polarization direction, i.e., $\theta=0, \pi$ and $\theta=\pi/2$. For $n_{\mathrm{d}}=18$, most angles achieve $P=1$.
But it's obvious that such symmetrical structure don't exist for $\phi=\pi/2$. Smaller-angle penetrabilities are enhanced, especially for $\theta=0$. While larger-angle penetrabilities are depressed.

\begin{figure}[!b]
\centering
\includegraphics[width=\linewidth]{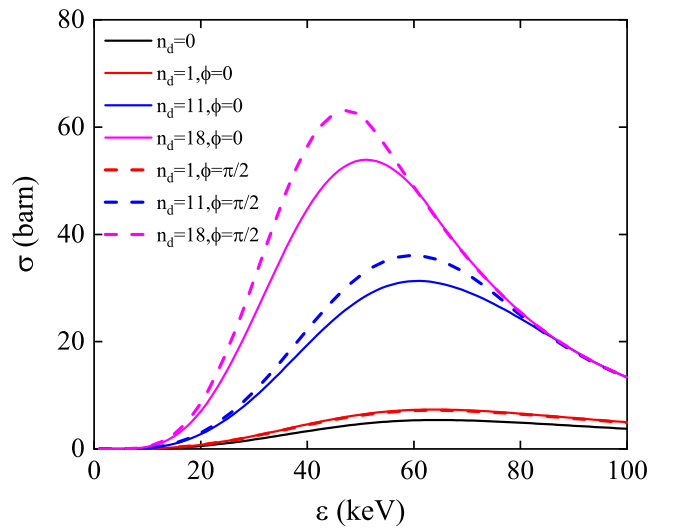}
\caption{(Color online) Fusion cross-sections versus collision energy for $n_{\mathrm{d}}=0,1,11$ and $18$.}
\label{figure4}
\end{figure}

The angle-averaged penetrability can be obtained by taking an average over the solid angle, that is,
\begin{eqnarray}
P_{\mathrm{ave}}\left(\varepsilon;\phi,n_{\mathrm{d}}\right)
&=&\frac{\int_{0}^{\pi} P\left(\theta;\phi,\varepsilon,n_{\mathrm{d}}\right) \sin\theta d\theta }{2}.
\label{Pave}
\end{eqnarray}

The penetrability versus the collision energy for different $n_{\mathrm{d}}$ values is shown in Fig. \ref{figure3} (b), indicating that the penetrability increases significantly with respect to the dimensionless parameter $n_{\mathrm{d}}$.

\subsection{DT fusion cross sections}
The nuclear fusion cross sections which are used to describe the probability that a nuclear reaction will occur are usually given in a phenomenological Gamow form as a product of three terms:
\begin{eqnarray}
\sigma\left(\varepsilon\right)=\frac{S\left(\varepsilon\right)}{\varepsilon}\exp\left(-\sqrt{\frac{\varepsilon_{\mathrm{G}}}{\varepsilon}}\right),
\label{Gamow}
\end{eqnarray}
where the term $1/\varepsilon$ is the geometrical cross sections, which is proportional to the square of the relative motion's de Broglie wavelength. And the term $\exp(-\sqrt{\varepsilon_{\mathrm{G}}/\varepsilon})$ is the tunneling probability through the Coulomb potential barrier, which holds as far as $\varepsilon\ll\varepsilon_{\mathrm{G}}$. For DT fusion, the Gamow energy factor is $\mathrm{\varepsilon_{\mathrm{G}}}=(e^{2}\sqrt{2m}/4\hbar\varepsilon_{0})^{2}=1.18\ \mathrm{MeV}$, so Eq. (\ref{Gamow}) applies very well to collision energies of $\varepsilon\leq 100\ \mathrm{keV}$. Astrophysical $S$ factor describes the nuclear physics within the nuclear potential effective range. In the absence of external electromagnetic fields, the $S$ factor can be given by a fitting function, which is the same as that in our previous paper \cite{Lv2019}.

\begin{figure}[!tb]
\centering
\includegraphics[width=\linewidth]{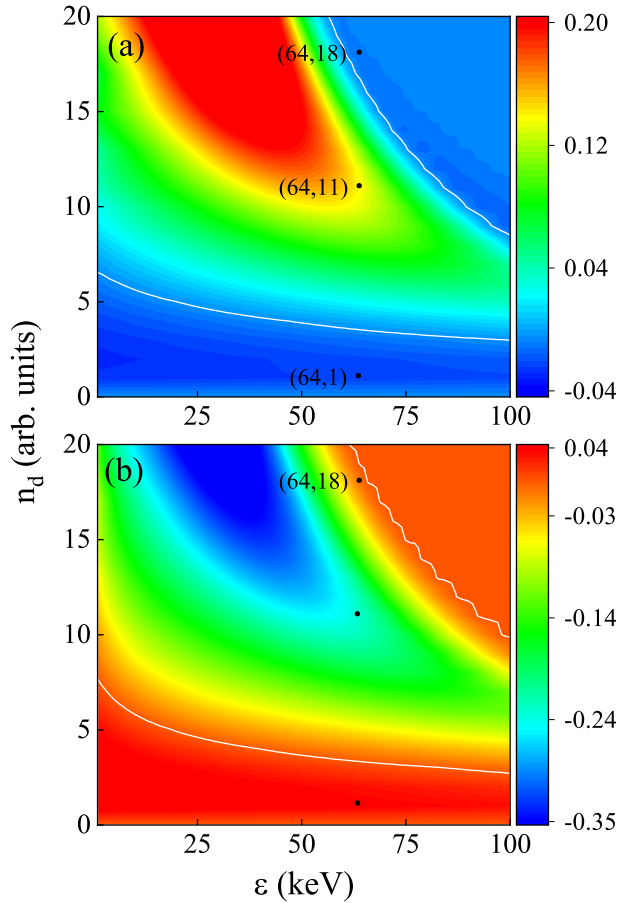}
\caption{(Color online) Contour plots of $\left(\sigma\left(\phi=\pi/2\right)-\sigma\left(\phi=0\right)\right)
/\sigma\left(\phi=0\right)$ for different $n_{\mathrm{d}}$. (a)$1\omega-2\omega$ and (b)$1\omega-3\omega$. }
\label{figure5}
\end{figure}

The laser intensities being chosen and regarded as very intense laser fields are still negligible compared to nuclear potentials. So we assumed that the laser fields do not alter the DT-fusion process from the fundamental level, such as affecting the astrophysical S function. Then we can get DT fusion cross sections according to Eq. (\ref{Gamow}) where the the tunneling probability can be calculate by angle-averaged penetrability Eq. (\ref{Pave}). The results are shown in Fig. \ref{figure4}. For $n_{d}=18$, the cross sections are shown to be enhanced by about $8$ times of magnitude for collision energy $\varepsilon=64$ keV and its maximum equals $63.05$ barns for $\varepsilon=47$ keV rather than $\varepsilon=64$ keV. This means that laser assisted deuteron-triton fusion can obtain bigger cross sections at smaller collision energy.

Contour plots of $\left(\sigma\left(\phi=\pi/2\right)-\sigma\left(\phi=0\right)\right)
/\sigma\left(\phi=0\right)$ for $n_{\mathrm{d}}\in[0,20]$ are shown in Fig. \ref{figure5}. Fig. \ref{figure5} (a) are $1\omega-2\omega$ fields. As shown in Fig. \ref{figure2} (a),(b),(A),(B), for $n_{\mathrm{d}}=1$, the potential difference for $\theta=\pi/2$ between $\phi=0$ and $\phi=\pi/2$ is very small, so the angle-averaged potential is primarily decided by other $\theta$ that the weighted coefficient are smaller. It's clear that $V_{\mathrm{eff}}(\phi=\pi/2$, $\theta=0)>V_{\mathrm{eff}}(\phi=0$, $\theta=0$ or $\theta=\pi)>V_{\mathrm{eff}}(\phi=\pi/2$, $\theta=\pi)$, Although the penetrability is exponentially dependent on the potential, our calculations show that the indices are very small, so the penetrability is not simply dependent on the smaller angle-averaged potential. We have shown that $P_{\mathrm{ave}}(\phi=0)>P_{\mathrm{ave}}(\phi=\pi/2)$ in Fig. \ref{figure3} (b).

For $n_{\mathrm{d}}=11$, the potential difference for $\theta=\pi/2$ between $\phi=0$ and $\phi=\pi/2$ becomes more and more clear, so a transition is expected. As shown in Fig. 2 (c),(d),(C),(D), for $\theta=\pi/2$, $V_{\mathrm{eff}}(\phi=0)>V_{\mathrm{eff}}(\phi=\pi/2)$, so we can see that $P_{\mathrm{ave}}(\phi=0)<P_{\mathrm{ave}}(\phi=\pi/2)$ for $\varepsilon=64$ keV in Fig. \ref{figure3} (b). This result demonstrates that penetrability is mainly decided by the extremum of the electric fields. For fundamental frequency and second-harmonic generation (SHG) laser, the minimum and the maximum of the electric fields occur at $\phi=0$ and $\phi=\pi/2$, respectively.

For $n_{\mathrm{d}}=18$, another transition occurs due to over-barrier effects, i.e., more and more angles achieve $P(\theta)=1$, as shown in Fig. \ref{figure2} (e),(f),(E),(F). For $\theta=\pi/2$, we can see that $V_{\mathrm{eff}}(\phi=0)$ and $V_{\mathrm{eff}}(\phi=\pi/2)$ are smaller than $\varepsilon=64$ keV. So the angle-averaged potential is primarily decided by other $\theta$ that the weighted coefficient are smaller. Here, smaller angles have higher peaks and broader tunneling widths for $\phi=\pi/2$, so we get $P_{\mathrm{ave}}(\phi=0)>P_{\mathrm{ave}}(\phi=\pi/2)$ in Fig. \ref{figure3} (b).
Fig. \ref{figure5} (b) are $1\omega-3\omega$ fields, which also have twice shifts of the peak value in the cross sections, but the phase relationship of the fusion cross-sections are opposite.

\section{Conclusion}
In summary, we show that DT fusion cross-sections can be enhanced depending on a dimensionless parameter $\mathrm{n_{d}}$ in the presence of high-intensity dichromatic laser fields with high-frequency limit. For $1\omega-2\omega$ fields, collision energy of $64$ keV and  $n_{\mathrm{d}}=18$, the fusion cross-section is approximately $8$ times as large, and the maximum fusion cross-sections can be enhanced to $63.05$ barns corresponding to $\varepsilon=47$ keV, which is approximately $15$ times that of the field-free case. In this situation we approximately estimate that the averaged fusion reactivity can be multiplied so that the Lawson criterion \cite{Lawson} might be reduced.

We find that there are twice shifts of the peak in the cross sections with the increase of $\mathrm{n_{d}}$. And different strong dichromatic laser fields(i.e. $1\omega-2\omega$ and $1\omega-3\omega$) have different shifts.

However, when calculating energy balance, we ignore the power that need to create and maintain the super-strong electromagnetic fields. This issue will be discussed in detail in future's work. On the other hand, this work focuses on the high-frequency limit. Extending these discussions to the situation of relatively low-frequency is undergoing.

\section*{Acknowledgments}
This work was supported by funding from China NSF No. 11775030 and NSAF No. U1930403.


\begin{thebibliography}{99}
\bibitem{Joachain}C. J. Joachain, N. J. Kylstra, R. M. Potvliege, Atoms in Intense Laser Fields. Cambridge University Press (2012).
\bibitem{Liu}J. Liu, Classical Trajectory Perspective of Atomic Ionization in Strong Laser Fields. Springer Berlin Heidelberg (2014).
\bibitem{Mima}K. Mima, J. Fuchs, T. Taguchi, et al., Matter and Radiation at Extremes \textbf{3}, 127 (2018).
\bibitem{Mangles}S. P. D. Mangles, C. D. Murphy, Z. Najmudin, et al., Nature \textbf{431}, 535 (2004).
\bibitem{Geddes}C. G. R. Geddes, Cs. Toth, J. van Tilborg, et al., Nature \textbf{431}, 538 (2004).
\bibitem{Faure}J. Faure, Y. Glinec, A. Pukhov, et al., Nature \textbf{431}, 541 (2004).
\bibitem{Karpeshin}F. F. Karpeshin, Physics of Particles and Nuclei \textbf{37}, 284 (2006).
\bibitem{Bai201801}D. Bai, D. M. Deng, Z. Z. Ren, Nuclear Physics A \textbf{976}, 23 (2018).
\bibitem{Qi2020}J. T. Qi, L. B. Fu, and X. Wang, Phys. Rev. C \textbf{102}, 064629 (2020).
\bibitem{Delion2017}D. S. Delion and S. A. Ghinescu, Phys. Rev. Lett. \textbf{119}, 202501 (2017).
\bibitem{Bai201802}D. Bai and Z. Z. Ren, Commun. Theor. Phys. \textbf{70}, 559 (2018).
\bibitem{Qi2019}J. T. Qi, T. Li, R. H. Xu, L. B. Fu, and X. Wang, Phys. Rev. C \textbf{99}, 044610 (2019).
\bibitem{Palffy}A. P$\acute{a}$lffy and S. V. Popruzhenko, Phys. Rev. Lett. \textbf{124}, 212505 (2020).
\bibitem{Ghinescu}S. A. Ghinescu and D. S. Delion, Phys. Rev. C \textbf{101}, 044304 (2020).
\bibitem{Friedemann2019}F. Queisser and R. Sch$\ddot{u}$tzhold, Phys. Rev. C \textbf{100}, 041601(R) (2019).
\bibitem{Lv2019}W. J. Lv, H. Duan, and J. Liu, Phys. Rev. C \textbf{100}, 064610 (2019).
\bibitem{Henneberger}W. C. Henneberger, Phys. Rev. Lett. \textbf{21} 838 (1968).
\bibitem{Wang2020}X. Wang, Phys. Rev. C \textbf{102} 011601(R) (2020).
\bibitem{Schafer 1992}K. J. Schafer, K. C. Kulander, Phys. Rev. A, \textbf{45}, 8026 (1992).
\bibitem{Chen Jing 1998}J. Chen, S. G. Chen, and J. Liu, CHINESE JOURNAL OF COMPUTATIONAL PHYSIC, \textbf{15}, 274 (1998).
\bibitem{Chen Jing 2000}J. Chen, J. Liu, and S. G. Chen, Phys. Rev. A, \textbf{61}, 033402 (2000).
\bibitem{Cheng Taiwang 1999}T. W. Chen, J. Liu, and S. G. Chen, Phys. Rev. A, \textbf{59}, 1451 (1999).
\bibitem{Cheng Taiwang 2000 1}T. W. Chen, J. Liu, and S. G. Chen, Commun. Theor. Phys. (Beijing, China), \textbf{33}, 33 (2000).
\bibitem{Cheng Taiwang 2000 2}T. W. Chen, J. Liu, and S. G. Chen, Phys. Rev. A, \textbf{62}, 033402 (2000).
\bibitem{Cheng Taiwang 2002}T. W. Cheng, X. F. Li, P. M. Fu, and S. G. Chen, Chinese Physics Letters, \textbf{19}, 1088 (2002).
\bibitem{Atzeni}Stefano Atzeni and J$\ddot{\mathrm{u}}$rgen Meyer-ter-Vehn, Inertial Fusion Beam plasma interaction, hydrodynamics, dense plasma physics. Clarendon press, Oxford (2003).
\bibitem{Gamow}G. Gamow, ZS. f. Phys. \textbf{51}, 204 (1928).
\bibitem{Bosch}H. S. Bosch and G. M. Hale, Nucl. Fusion \textbf{32}, 611 (1992).
\bibitem{Landau}L. D. Landau and E. M. Lifshitz, Quantum Mechanics Non-relativistic Theory. Pergamon Press, London-Paris (1958).
\bibitem{Lawson}J. D. Lawson, Proceedings of the Physical Society (London) \textbf{70}, 6 (1957).

\end{thebibliography}
\end{document}